\documentclass[utf8]{FrontiersinHarvard} 

\usepackage{url,hyperref,lineno,microtype,subcaption}
\usepackage[onehalfspacing]{setspace}

\def\keyFont{\fontsize{8}{11}\helveticabold }
\def\firstAuthorLast{Hackney {et~al.}}

\def\Authors{
Kai Alexander Hackney\,$^{1}$,
Lucas Guarenti Zangari\,$^{1}$,
Jhonathan Sora-Cardenas\,$^{1,\dagger}$,
Emmanuel Munoz\,$^{1,\dagger}$,
Sterling R. Kalogeras\,$^{1,\dagger}$,
Betsy DiSalvo\,$^{1}$,
Pedro Guillermo Feijóo-García\,$^{1,*}$
}

\def\Address{
$^{1}$College of Computing, Georgia Institute of Technology, Atlanta, GA, USA
}

\def\corrAuthor{Pedro Guillermo Feijóo-García}

\def\corrEmail{pfeijoogarcia@gatech.edu}

\begin{document}
\onecolumn
\firstpage{1}

\title[Voice, Personality, and Gender in Human-Agent Interaction]{Exploring the Interplay Between Voice, Personality, and Gender in Human-Agent Interactions} 

\author[\firstAuthorLast ]{\Authors} 
\address{} 
\correspondance{} 

\extraAuth{}

\maketitle

\begin{abstract}

\section{}
To foster effective human-agent interactions, designers must understand how vocal cues influence the perception of agent personality and the role of user-agent alignment in shaping these perceptions. In this work, we examine whether users can perceive extroversion in voice-only artificial agents and how perceived personality relates to user-agent synchrony. We conducted an exploratory study with 388 participants, who evaluated four synthetic voices derived from human recordings, varying by gender (male, female) and personality expression (introverted, extroverted). Our findings suggest that participants were able to differentiate perceived extroversion across the female voice conditions, but not consistently across the male voice conditions. We also observed preliminary evidence consistent with perceived personality synchrony, particularly in participants' evaluations of the first agent encountered. These findings represent an initial step toward understanding personality synchrony in voice-based human-agent interactions. Because each experimental condition was represented by a single synthesized voice and no objective acoustic validation of the intended personality manipulation was conducted, these findings should be interpreted within the exploratory scope of this study. We discuss the implications of these findings while considering the limitations of stimulus diversity and voice representation, and outline implications for the design of voice-based agents, particularly regarding the interaction between gender, personality perception, and initial user impressions. This paper contributes exploratory evidence and methodological insights for investigating the interplay of user-agent personality and gender synchrony in the design of human-agent interactions.

\tiny
 \keyFont{ \section{Keywords:} Human-AI Interaction, Human-Centered AI, Similarity-Attraction, User-Agent Synchrony} 
\end{abstract}

\section{Introduction}

Human-agent interactions are increasingly mediated through voice-based interfaces, where vocal cues play a central role in shaping how users perceive and relate to artificial agents. Prior work has shown that users tend to favor agents that resemble them, a phenomenon known as the similarity-attraction effect \cite{byrne1961interpersonal}, which extends to human-agent interactions \cite{zhou2019trusting, braun2019your, bernier2010similarity}. In this space, personality alignment between users and agents has been identified as a key factor shaping user preference and trust \cite{maxim2023impact}. However, much of this work has examined personality through a limited set of vocal features, or in combination with non-vocal cues such as visual embodiment or behavior \cite{cumbal2024let, ishii2020impact, dollinger2024exploring}.

Despite these advances, we still have a limited understanding of how users perceive personality in voice-only agents, where no visual or embodied cues are present. This becomes particularly important as voice-based systems continue to expand in real-world applications. Prior research has also suggested that similarity-attraction effects are more strongly driven by perceived personality alignment than by surface-level vocal similarity \cite{nass2000does}. This naturally raises the question of whether users can meaningfully perceive personality, specifically extroversion, from voice alone, and how this perception interacts with user characteristics such as gender and self-reported personality.

In this paper, we present an exploratory study examining how perceived extroversion and gender cues in voice-only agents shape user perceptions in human-agent interactions. We focus specifically on artificial voices without visual embodiment. To do so, we generated synthetic agent voices from human recordings, selecting four voice samples based on gender (male, female) and extroversion levels (introverted, extroverted). Thus, this study is intended as an exploratory investigation of how participants perceived these synthesized voice conditions. Participants (n = 388) evaluated these voices in a controlled setting, where each participant interacted with two agents of the same gender but differing in extroversion. 

This design allows us to explore both whether users can differentiate perceived extroversion across the synthesized agent voice conditions and how user-agent personality synchrony influences these perceptions. Based on this, we address the following research questions:

\begin{itemize}
\item \textbf{RQ1 - {\textit{Perceived Extroversion in Voice}}:} To what extent can users differentiate perceived extroversion in voice-only agents, and how does this vary across gender?
\item \textbf{RQ2 - {\textit{User-Agent Personality Synchrony}}:} To what extent do users project their own personality traits onto virtual agents, and how does perceived personality similarity influence user evaluations?
\end{itemize}

We assess personality using the Ten Item Personality Measure (TIPI) \cite{GOSLING2003504}. Given the scope of this work, we focus specifically on the extroversion dimension and interpret our findings in terms of perceived extroversion in agent voices.

\section{Background}

The influence of a personality similarity-attraction effect on people's trust and perception has been a central focus in human-agent interaction studies. \cite{zhou2019trusting} found that agents that align their personality traits with user expectations are trusted more, demonstrating the role of personality congruence. In the setting of in-car speech interfaces, \cite{braun2019your} showed that there was higher likability and trust for an assistant who matched the personality of the driver, and lower likability and trust for incorrectly matched personalities. \cite{nass2000does} confirmed that people demonstrated a similarity-attraction effect regarding personality for computer-generated speech, even when personality is not relevant to that speech. In a study by \cite{maxim2023impact}, it was found that matching a virtual human's vocalic extroversion level with a participant's extroversion level did not result in a similarity-attraction effect or impact the effectiveness of the mental health intervention. Instead, the participant's own extroversion level was the primary factor influencing their ratings. \cite{cumbal2024let} looked at three situations when an agent got interrupted by a human. The agent would either ignore the interruption, yield to the human, or acknowledge the interruption and finish its thought. The study's results showed that this change in behavior impacted people's perceptions of the agent's personality as well as how much they enjoyed interacting with the agent.

Other research has focused on methods for detecting and interpreting personality in both agents and users. \cite{ishii2020impact} showed that Big Five personality traits influence the automatic generation of nonverbal behaviors in virtual agents, showing that combining spoken language features with personality information enables the creation of more natural and socially believable behaviors such as eye gaze, head nods, hand gestures, and upper-body posture. \cite{berkovsky2019detecting} explored using eye-tracking data to infer personality traits of study participants. They found that eye movement patterns correlate with and can predict personality traits, suggesting that personality inference can allow agents to dynamically adapt and present personalities in ways that foster trust and alignment. \cite{peltonen2020phones} discovered that smartphone usage patterns provide predictive signals of personality. \cite{dollinger2024exploring} did a study evaluating how participants evaluate an agent's personality and self-relatability based on their interaction. They found a positive relationship between personality similarity, self-attribution, and a sense of behavior-similarity.

Even for agents with a dissimilar appearance, the sensation of appearance similarity is conserved. \cite{lee2019does} showed that individuals either perceive the conversational agent as a human, speaker, system, or a space object. Participants who perceived the agent as human tended to base their gender decision on the agent's vocal tone. Participants who perceived the agent as a speaker personified the physical device that they used to interact with the agent. Participants who perceived the agent as a system tended to visualize it as a machine. Finally, those who perceived the agent as a space object oftentimes reflected it as an open and shared resource. \cite{esterwood2021meta} analyzed 26 other HRI studies on if human personality can accurately predict robot acceptance. Their results did find a positive correlation, but they also found gaps in the literature, such as the need to examine a larger age range of participants.

\subsection{Gender Cues and Virtual Agents}
Previous studies have investigated the influence of gender on users' relationships with virtual humans. \cite{feijoo2021effects} show that gender and vocal cues, such as accent, shape how users evaluate agents, with biases influencing whether they are perceived as credible or expert, highlighting that such perceptions are socially and culturally conditioned and making gendered and vocal signals critical factors in agent design. Similarly, \cite{gilpin2018perception} found that pitch-based cues predicted openness and neuroticism with particular accuracy, with conscientiousness and agreeableness more reliably classified in female speakers and neuroticism more accurately detected in male speakers, demonstrating that vocal features can act as gender-specific markers of personality that conversational agents may leverage to enhance social alignment and believability.
The similarity-attraction effect has been widely documented in psychology and HRI, showing that people tend to perceive agents as more friendly when they share similar preferences. \cite{bernier2010similarity} demonstrated this principle in human-robot interactions, suggesting that general social mechanisms such as similarity-attraction also apply to virtual agents, shaping responses to cues like gender and voice. In line with this, \cite{feijoo2024exploring} examined how gender synchrony between users, virtual agents, and designers influences trust and intentions toward wellness practices, providing evidence that demographic cues such as gender and vocal traits significantly shape perceptions of credibility and rapport, while highlighting user-agent synchrony as a mechanism for improving acceptance and effectiveness in mental health contexts.

Recent work suggests that gender similarity interacts with other design factors such as voice and appearance. \cite{feijoo2024effects} investigated user-agent-designer interactions in the context of mental health support, examining whether demographic similarities influenced engagement. Their findings show that both user-agent and designer-agent similarity significantly shape how virtual humans are created and received, with visual and auditory cues often exerting greater influence than verbal information about the designer. Overall, these results indicate that gender functions as an important social cue in shaping impressions of credibility and expertise, but its impact is moderated by other factors such as voice, appearance, and contextual framing. Thus, while gender similarity can enhance acceptance and rapport, its contribution is best understood in relation to the broader set of design features that collectively influence user trust and engagement.

\subsection{Vocal Cues and Virtual Agents}
Many studies refer to vocal cues as a fundamental attribute that significantly affects the human-agent interaction. Research has revealed how, for humans, certain vocal attributes correlate to certain personalities, such as faster response times in extroverted individuals compared to introverted individuals, as well as the loudness of individuals regarding their personality \cite{lee2021speech}. \cite{gilpin2018perception} identified which vocal cues revealed personality traits through support vector machine classifiers. The subgroups mostly affected by pitch were openness and neuroticism, while prediction accuracy for conscientiousness and agreeableness were higher in the female subgroup, while neuroticism had improved prediction accuracies in the male subgroup. The results found in humans serve as the basis for the findings in virtual agents.

Human-agent interactions are significantly influenced by the agent's vocal cues. When presented in a real-world setting, users perceived their interactions as much more realistic and natural with generative AI voices rather than traditional text-to-speech, primarily due to the vocal attributes of the artificially generated voice compared to the monotony of traditional text-to-speech implementations \cite{maxim2025perceived}. The irregularities and variety in vocal cues directly affect how a user will perceive the human-agent interaction, as it parallels social cues, enhancing the believability of the interaction \cite{ishii2020impact}. Looking at previous studies, \cite{nass2000does} observed whether people demonstrated a similarity-attraction effect regarding personality for computer-generated speech, and given the experiment, the results show how the user's personality made the most difference in the effect, rather than the individual's voice. Similarly, \cite{maxim2023impact} was able to notice a similar effect when measuring vocalic extroversion level, as the participant's own extroversion level made the most difference. However, it is important to note that personality is heavily integrated in their voice through their vocal cues, and it is important to keep this in mind when designing virtual agents. Across these various studies, it is clear as to how an agent's vocal cues had a great effect in the human-agent interactions.

Individuals can also develop a bias based on an agent's vocal traits. \cite{braun2019your} designed in-car speech interfaces and tested various agents, and their results show how drivers enjoyed and trusted agents that better aligned with their own personality than those with a different personality based on their interactions while driving. These findings reveal how tuned human ears are in identifying and distinguishing a personality through a primarily voice-only modality, and are able to capture and perceive an identity. Similarly, \cite{zalake2021effects} reveals how vocal and verbal strategies used by agents can affect user decisions, thus showing how vocal cues influence behavioral outcomes. Looking at agents another way, \cite{lee2019does} demonstrates how individuals conceptualize and visualize various voice-only agents based on their vocal cues, and those who illustrated a human-like figure often based their decision on certain vocal traits and compared that to individuals they have previously experienced. Given certain vocal cues, it is clear to see how individuals can be biased towards some vocalic characteristics over others.

\subsection{User-Agent Similarity in Human-Agent Interactions}
Synchrony, which refers to the alignment of an agent with a user in terms of cues, styles, and behavior, forms a key piece of human-agent interaction research. Furthermore, synchrony plays a critical role in developing trust, credibility, and a user's willingness to interact with virtual agents \cite{you2022does, lucas2014s}.

Previous work dives into providing a road-map for eliciting user-trust and self-disclosure through communicative alignment. Such work stresses the importance of consistency, transparency, and socially attuned responses as valuable instruments for promoting user trust \cite{you2022does}. Culminating with this are the findings of other previous work showing that humans often entrust private information to virtual humans over actual humans due to social evaluation, judging, and being mitigated through the virtual human context \cite{lucas2014s}. Together, these findings suggest increases of openness in sensitive domains/conversations reflect the careful tuning of synchrony and alignment in agent communication, ultimately fostering safe spaces. 

However, when viewing synchrony under demographic and social dimensions, it becomes evident that trust and credibility hinge not only on user-agent similarity cues but on a combination of user-designer and user-agent similarity \cite{feijoo2024exploring}. This yields a three-way intersection between users, agents, and designers that, according to previous work, leads to increases in mental health practices when demographic factors such as gender and age are aligned \cite{feijoo2024exploring}. These effects and their respective narratives showcase a new view of synchrony, one shaped not only by gender but also impacted by age-proximity. These factors, some superficial and others less, reveal that shaping credibility in virtual humans is a more complex balance than previous work has thought. 

The tightrope act synchrony factors play has also been the focus of previous work. When examining human-robot interactions, prior research highlights that humans perceive robots more favorably when there are shared characteristics and preferences between the two \cite{bernier2010similarity}. Said work stresses that synchrony is not merely an aesthetic factor but rather one that taps into pre-established human-social evaluation mechanisms. Interestingly, synchrony between user and agent personality traits enhances relatability and self-attribution in psychotherapy contexts, despite differences in user and agent appearances \cite{dollinger2024exploring}. 

In the end, previous work agrees that user-agent synchrony is a mix of several factors, including communicative behavior, demographic cues, and personality similarity. Together, these three factors comprise a powerful driver for credibility and engagement in virtual humans. It's clear that designing virtual humans matters not merely on effective communication but also on effective social and psychological similarity between users and agents.

\section{Methodology}

The user study took place in Spring 2025 at a large Southeastern institution in the United States. To support the generation of the artificial voices used in the study, we conducted a separate preliminary data collection with online participants in the United States. In this preliminary phase, participants were assessed on personality traits and demographics, and submitted voice recordings reading standardized passages \cite{van1996speech, fairbanks1960voice}. The goal of this phase was not to build a broad or representative dataset, but to obtain controlled voice samples aligned with gender and extroversion levels for use in the main study. Hence, the resulting synthesized voices were intended to support an exploratory examination of personality perception rather than to represent the full range of vocal variation within each experimental condition.

Both the preliminary data collection and the main study were approved by the Institutional Review Board of the Georgia Institute of Technology (Protocol No. IRB2025-303). The main study consisted of one week of asynchronous, fully digital data collection conducted via institutional email and Qualtrics.

\subsection{Preliminary Data Collection: Participants}

The preliminary data collection aimed to obtain voice samples aligned with extreme values of extroversion across gender categories. We recruited adults (18 years of age or older) through the crowd worker platform \cite{Prolific2025}, with inclusion criteria requiring participants to be fluent in English and located in the United States.

A total of 346 participants responded to the study, of whom 13 submitted voice recordings using the \cite{Phonic2025} platform. The low submission rate reflected the optional nature of the voice-recording task, which was conducted as a separate phase to obtain candidate recordings for stimulus generation. After excluding two recordings that did not meet the predefined quality criteria (i.e., audio quality and adherence to the standardized reading script), 11 usable voice recordings remained. These recordings were not intended to provide broad or representative coverage of vocal diversity, but rather to identify voice samples at the extremes of the extroversion dimension for the generation of the synthesized voices used in the main exploratory study.

Among these participants, ages ranged from 20 to 61 years, and gender distribution included 6 male and 5 female participants. All participants were fluent in English, with one participant also reporting frequent use of Hindi.

\subsection{Preliminary Data Collection: Data Analysis}

The goal of this phase was to identify voice samples corresponding to extreme values of extroversion within each gender category. Based on participants' responses to the TIPI scale \cite{GOSLING2003504}, we selected one participant per condition (male introverted, male extroverted, female introverted, female extroverted) using the highest and lowest extroversion scores available within each group.

Although the TIPI was originally developed as a self-report measure, in this study we use its extroversion items as a structured framework for capturing participants' perceptions of the synthesized voices. Our objective is not to assess the "true" personality of the agents, but rather to quantify how participants interpreted the personality conveyed by each voice, allowing direct comparisons between self-reported and perceived extroversion under controlled experimental conditions.

For male participants, one individual scored the minimum value (1) and one the maximum value (7) on extroversion. For female participants, one individual scored the maximum value (7), while the lowest available score was 2.35. Selected recordings were manually reviewed to ensure audio quality and adherence to the provided script before being used to generate the synthesized voices employed in the main exploratory study.

Consistent with the exploratory nature of this work, each experimental condition was represented by a single synthesized voice derived from one selected recording. The resulting stimuli were intended to provide controlled experimental conditions rather than representative examples of vocal personality expression within each gender and extroversion category. Consequently, any observed differences between conditions cannot be unequivocally attributed to the intended personality manipulation, as they may also reflect stimulus-specific vocal properties, such as accent, speaking style, or other sociophonetic characteristics. Moreover, the synthesized voices were not objectively evaluated to verify that the intended personality manipulation was reflected in their acoustic characteristics. These considerations should be taken into account when interpreting our findings and are discussed further in Sections \ref{results} and \ref{limitations}.

\subsection{Preliminary Data Collection: Voice Generation} \label{generation}

After selecting the recordings corresponding to each condition, we used ElevenLabs \cite{ElevenLabs2025} to generate synthetic voices for the main study. All voices were generated using the same model (Eleven Multilingual V2) and identical parameter settings (\textit{Stability}: 50\%, \textit{Similarity}: 75\%, \textit{Style Exaggeration}: 0\%, \textit{Speaker boost}: on). This ensured consistency across voices, with the only variation introduced through the input recordings associated with each gender and extroversion condition.

The use of a single synthesis pipeline and fixed parameters was intended to reduce variability introduced by the generation process. However, since each condition was seeded from a single human recording, differences across voices may still reflect idiosyncratic vocal characteristics in addition to perceived personality cues. Therefore, the synthesized voices should be interpreted as controlled experimental stimuli for this exploratory investigation rather than representative exemplars of vocal personality expression.

All generated voices read the same script, adapted from Job Burnout \cite{annurev:/content/journals/10.1146/annurev.psych.52.1.397}, to ensure consistency in linguistic content. Although the source material discusses job burnout, the passage was selected because it provides a continuous, standardized narrative suitable for voice synthesis while minimizing abrupt changes in sentence structure, emotional expression, and speaking style across conditions. The script therefore served as a standardized reading passage rather than as content intended to influence participants' personality judgments and lasted an average of four minutes and 42 seconds across conditions.

\subsection{Participants}

A total of 429 adult (18 years of age or older) participants were recruited at a large Southeastern institution in the United States. Of these, 388 participants completed the survey and provided informed consent, and are included in our analysis. Participants were enrolled in the Spring of 2025 and recruited from a single course within the College of Computing at the Georgia Institute of Technology.

While our sample size exceeds commonly used thresholds for statistical analyses, we note that recruitment was based on course enrollment rather than a priori power analysis. Furthermore, because participants were recruited from a single computing course at one institution, the study population reflects a relatively homogeneous academic context. Hence, our findings should be interpreted within the context of this sampling approach and should not be assumed to generalize to broader populations.

Participants reported ages of 18 years (n = 114), 19 years (n = 142), 20 years (n = 93), 21 years (n = 29), 22 years (n = 4), and 23--29 years (n = 6). Participants reported spoken languages indicating that 157 were monolingual and 231 multilingual. Participants reported being from 28 different countries, including the United States (n = 278) and other countries (n = 110). Gender distribution was male (n = 276), female (n = 102), and other or not reported (n = 10). The resulting sample was predominantly composed of young adults (18 to 20 years old) and male participants, characteristics that should be considered when interpreting subgroup comparisons throughout this exploratory study.

\subsection{Intervention}

This study was conducted online and asynchronously via Qualtrics. The design follows a mixed structure, combining a within-subject component and between-subject factors. Each participant evaluated two voices (within-subject factor: extroversion), while group assignment determined the gender of the voices and their presentation order (between-subject factors).

\begin{figure}[th]
 \centering
 \includegraphics[scale=0.5]{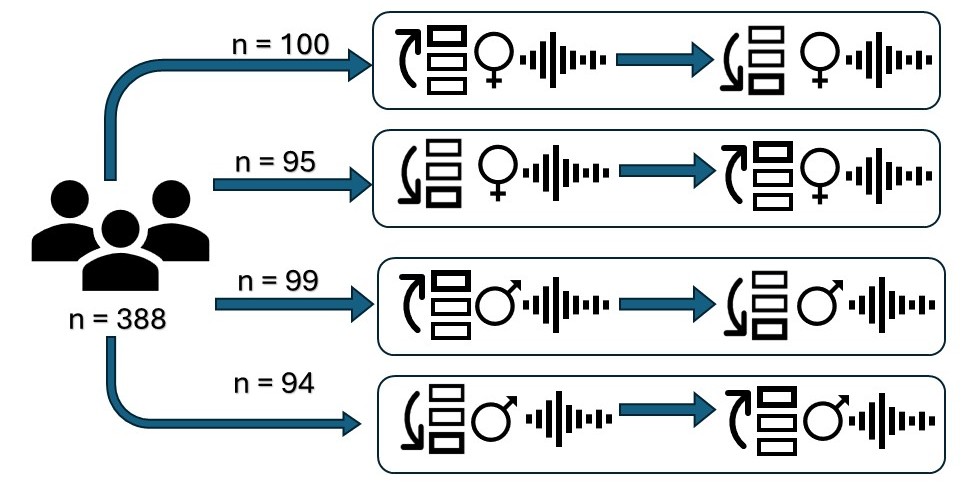}
 \caption{Participant assignment across the four intervention groups. Each participant evaluated two synthesized voices of the same gender, with groups varying by voice gender and presentation order of the introverted and extroverted conditions.}
 \label{fig:Protocol}
\end{figure}

Participants were randomly assigned to one of four groups (see Figure \ref{fig:Protocol}), which determined both the gender of the synthesized voices as well as whether participants encountered the extroverted or introverted voice first:

1) Extroverted Male followed by Introverted Male  
2) Introverted Male followed by Extroverted Male  
3) Extroverted Female followed by Introverted Female  
4) Introverted Female followed by Extroverted Female  

Each participant listened to two voices of the same gender, differing in extroversion. To ensure consistency in linguistic content, all synthesized voices read the same standardized script (see Section \ref{generation}). Participants listened to each recording in its entirety before completing the corresponding questionnaire. This design allows us to examine perceived extroversion differences within participants, as well as the effects of gender and presentation order across groups.

The estimated completion time for the study was 20 minutes. While recorded timestamps indicate longer durations in some cases, these are likely due to participants not completing the survey in a single sitting.

\subsection{Analysis Strategy}\label{analysis}

Participants' responses to the TIPI questionnaire \cite{GOSLING2003504} were used to compute extroversion scores for both self-assessment (Block 2) and perceived agent personality (Blocks 3a and 3b). This allowed us to derive, for each participant, an extroversion score for themselves and for each of the two voices they evaluated.

Given the mixed structure of the study, our analyses explores (1) perceived extroversion across voice conditions, and (2) the relationship between participant personality and perceived agent personality. We consider both within-participant comparisons (differences between the two voices evaluated by the same participant) and between-group comparisons (differences across gender and presentation order conditions). Although this study is exploratory, inferential statistical analyses were conducted to facilitate systematic comparisons across the synthesized voice conditions and to identify patterns that may inform future confirmatory studies. Thus, these analyses are interpreted within the context of the synthesized voice conditions examined in this study and should not be taken as definitive evidence of broader personality perception effects.

Because the extroversion scores are derived from Likert-type responses and treated as ordinal, we employed non-parametric inferential statistical methods \cite{sullivan2013analyzing}. Specifically, we used Mann-Whitney U tests \cite{mcknight2010mann, mann1947test} for between-group comparisons and paired comparisons within participants where appropriate. To examine associations between participant personality and perceived agent personality, we used Spearman's correlation \cite{de2016comparing}.

To mitigate the risk of inflated Type I error due to multiple comparisons, we applied appropriate corrections (i.e., the Holm-Bonferroni method \cite{holm1979simple})  when interpreting statistical significance. All analyses were conducted using a significance threshold of $\alpha = 0.05$. These analyses are therefore interpreted as exploratory comparisons intended to identify patterns within the synthesized voice conditions examined rather than to establish broadly generalizable effects or causal evidence regarding vocal personality perceptions.

\section{Findings and Results}\label{results}

\subsection{User Perceptions and the Interplay of Gender and Personality}
A total of 388 participants completed the task. We examined participants' ability to distinguish \textit{perceived extroversion} from voice-only agents. Observed patterns differed across the female and male voice conditions.

For female voices, participants (n=195) significantly differentiated between the introverted and extroverted voice conditions. We observed a significant negative correlation (Spearman's $r_s=-0.533$, $n=195$, $p < .001$), indicating that participants consistently perceived differences in extroversion across female voices. In contrast, for male voices, participants (n=193) did not show a significant association between perceived extroversion and the intended personality condition ($r_s=0.0083$, $n=193$, $p = 0.30$), suggesting that participants were not able to reliably distinguish extroversion levels across the male voice conditions.

\begin{figure}[th]
 \centering
 \includegraphics[scale=0.5]{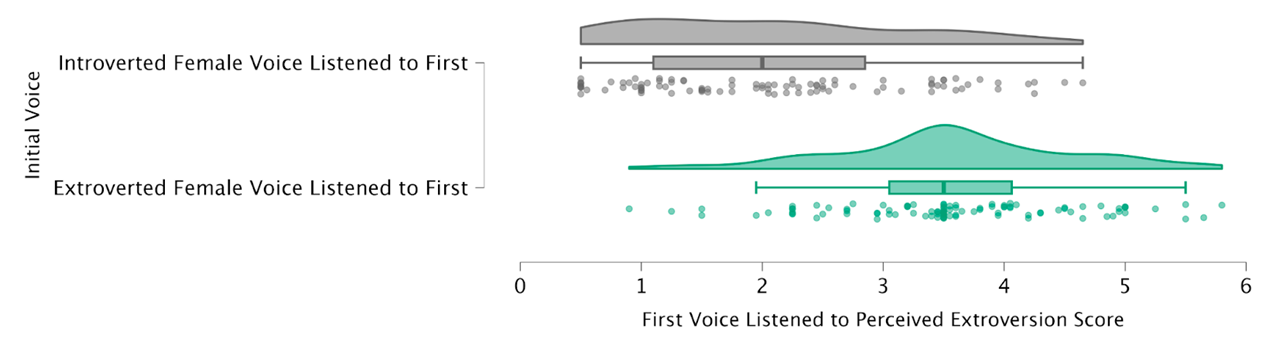}
 \caption{Extroversion Score Distribution (TIPI \cite{GOSLING2003504}) for Users' First Interaction with a Female Voice}
 \label{fig:Female Voices}
\end{figure}

To examine whether order of exposure influenced perceived extroversion, we conducted between-subject comparisons based on the first voice encountered. Given the ordinal nature of the TIPI-based extroversion scores and the lack of normality assumptions, we employed Mann-Whitney U tests for these comparisons.

\begin{figure}[th]
 \centering
 \includegraphics[scale=0.5]{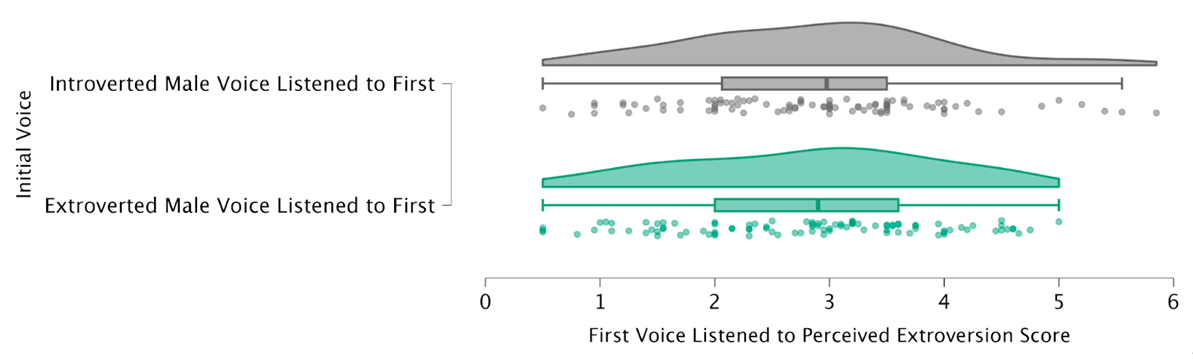}
 \caption{Extroversion Score Distribution (TIPI \cite{GOSLING2003504}) for Users' First Interaction with a Male Voice}
 \label{fig:Male Voices}
\end{figure}

For the female agent, participants who listened to the introverted voice first (n=95) were compared with those who listened to the extroverted voice first (n=100). To further investigate RQ1, we compared perceived extroversion scores between the two female voice conditions. We observed a significant difference in perceived extroversion between conditions (Mann-Whitney U test, $U=7915$, $p < .001$; see Figure \ref{fig:Female Voices}), indicating that participants differentiated the two female voice conditions in terms of perceived extroversion.

For the male agent, participants who listened to the introverted voice first (n=94) were compared with those who listened to the extroverted voice first (n=99). To further investigate RQ1, we compared perceived extroversion scores between the two male voice conditions. No significant difference was observed (Mann-Whitney U test, $U=4576$, $p = 0.85$; see Figure \ref{fig:Male Voices}), suggesting that participants did not reliably differentiate the two male voice conditions in terms of perceived extroversion.

\textbf{RQ1 - \textit{Perceived Extroversion in Voice}:} To what extent can users differentiate perceived extroversion in voice-only agents, and how does this vary across gender?

Our findings indicate that participants differentiated perceived extroversion across the female voice conditions, whereas no such differentiation was observed for the male voice conditions. This pattern was consistently observed across the exploratory analyses conducted in this study, including comparisons based on participants' first exposure to a voice. These findings suggest that the intended extroversion difference was more readily differentiated within the female voice conditions than within the male voice conditions examined in this study.

In contrast, the male voice conditions were not reliably differentiated in terms of perceived extroversion. Despite being generated using the same procedure, perceived extroversion scores for male voices remained statistically indistinguishable between the two male voice conditions.

Several factors may account for this discrepancy. First, the mapping between self-reported extroversion and vocal expression may not have translated equivalently across the selected voice samples. Second, uncontrolled vocal characteristics (e.g., accent, prosody) may have introduced variability unrelated to the intended personality manipulation. For example, the introverted male voice exhibited a noticeable Southern U.S. accent, which may have influenced participants' perceptions independently of extroversion. Because each experimental condition was represented by a single synthesized voice, the present data do not allow these alternative explanations to be disentangled.

Given that each condition was represented by a single synthesized voice, these findings should be interpreted with caution. The observed differences may reflect not only personality cues but also idiosyncratic vocal characteristics. Future work should incorporate a larger and more controlled set of voice stimuli, including systematic variation of acoustic features, to disentangle personality expression from other sociophonetic factors.

Overall, the present findings reveal an asymmetry between the female and male voice conditions examined in this exploratory study. Whether this asymmetry reflects differences in the expression of extroversion, stimulus-specific vocal characteristics, or a combination of both cannot be determined from the present study and should be examined in future work.

\subsection{User Perceptions and User-Agent Personality Synchrony}
We also explored patterns of personality synchrony between participants and the perceived personality of the first voice they interacted with. Across the full sample, we initially observed a positive correlation ($r_s=0.124$, $n=388$, $p=0.025$), suggesting a tendency for participants to perceive the first agent as having a personality similar to their own. However, after applying Holm-Bonferroni corrections \cite{holm1979simple} for multiple comparisons, this effect did not remain statistically significant (adjusted $\alpha = 0.017$).

When disaggregating by participant gender, a significant correlation was observed among male participants ($r_s=0.150$, $n=276$, $p=0.013$), which remained significant after correction. In contrast, no significant correlation was observed among female participants ($r_s=0.025$, $n=102$, $p=0.80$). Because the female participant subgroup was substantially smaller than the male subgroup, the absence of a significant association should not necessarily be interpreted as evidence that such an effect is absent.

Taken together, these findings provide preliminary evidence consistent with personality synchrony, although this evidence was not uniform across analyses. While the overall correlation across the full sample did not remain statistically significant after applying Holm-Bonferroni corrections \cite{holm1979simple}, the effect observed among male participants remained significant. Therefore, the observed relationship should be interpreted as an exploratory finding rather than evidence of a general synchrony effect across participants.

These observations indicate that the relationship between self-perceived personality and perceived agent personality varied across the participant groups examined in this exploratory study. Specifically, synchrony effects were observed only in the first interaction and primarily among male participants. Whether this pattern reflects genuine differences across participant groups or is influenced by the unequal subgroup sizes cannot be determined from the present study and should be investigated in future work.

In response to \textbf{RQ2 - \textit{User-Agent Personality Synchrony}}: \textit{To what extent do users project their own personality traits onto virtual agents, and how does perceived personality similarity influence user evaluations?}, our findings provide preliminary evidence consistent with the possibility that users may project their own personality onto agents during initial exposure. Within this study, the observed pattern was limited to participants' first interaction with a synthesized voice and to the male participant subgroup.

Importantly, when participants evaluated the second agent, no significant relationship was observed between participants' self-reported extroversion and the perceived extroversion of that agent. Instead, a significant relationship emerged between the perceived extroversion of the first and second agents, suggesting that participants relied on their initial exposure as a reference point. This pattern is consistent with an anchoring effect, where early impressions shape subsequent evaluations.

Together, these findings suggest that perceived personality synchrony may occur primarily in early interactions, when users have limited information about an agent. As additional information becomes available, user perceptions appear to shift from self-referential judgments to comparative evaluations based on prior interactions.

While these results point to a potential role of user characteristics in shaping early perceptions of agent personality, the evidence is not uniform across the full sample. Therefore, conclusions regarding user-agent personality synchrony should be interpreted cautiously. Additionally, because each experimental condition was represented by a single synthesized voice, it remains unclear whether the observed synchrony reflects participants' responses to the intended personality manipulation or to stimulus-specific vocal characteristics. Future work should further investigate these effects using larger and more diverse samples, more balanced participant groups, as well as more rigorously controlled experimental manipulations of agent personality cues and multiple synthesized voice stimuli per experimental condition, to better understand when and how synchrony emerges in human-agent interaction contexts such as education or mental health support.

\section{Conclusions and Future Work}

This exploratory study investigated how agent gender and voice-based personality cues influence users' perceptions of virtual agents. Our findings suggest that participants differentiated perceived extroversion across the female voice conditions, whereas no such differentiation was observed across the male voice conditions. These results suggest that, under the current experimental conditions, the intended extroversion difference was more readily differentiated and more readily differentiated within the female voice conditions than within the male voice conditions examined in this study.

In addition to differences across agent gender, we examined patterns of user-agent personality synchrony. We found preliminary evidence consistent with the possibility that participants may project their own personality onto a voice-based agent during initial interactions. However, this effect was not observed across the full sample after correcting for multiple comparisons. The observed relationship remained significant among male participants, although whether this reflects genuine subgroup differences or the unequal participant distribution remains an open question. Furthermore, synchrony effects were limited to the first interaction. By the second evaluation, participants no longer exhibited evidence consistent with personality synchrony, suggesting that additional factors may have influenced subsequent judgments. Further research is needed to fully understand this phenomenon.

Taken together, these findings highlight both the potential and the limitations of voice-based cues in shaping perceived personality in human-agent interactions. Because each experimental condition was represented by a single synthesized voice, our findings should be interpreted as exploratory evidence derived from the specific voice conditions examined in this study. Although voice can convey personality traits in some contexts, our observations suggest that the perception of these traits may depend not only on user characteristics and the stage of interaction, but also on stimulus-specific vocal characteristics.

Future work should extend these findings by incorporating a larger and more diverse set of voice stimuli, with greater control and objective validation of acoustic features such as pitch, tempo, and prosody. Increasing stimulus diversity would help disentangle personality-related cues from idiosyncratic vocal characteristics. Additionally, future research should further investigate the conditions under which user-agent personality synchrony emerges, including its dependence on user traits, interaction context, and information availability. Exploring these dynamics in applied settings, such as education and mental health support, may provide further insight into how personality cues can be effectively designed to support user engagement and interaction outcomes.

Finally, despite its exploratory nature, this study provides preliminary design considerations for voice-based agents. Designers should consider how personality cues are introduced during early interactions, as initial impressions may shape subsequent perceptions. At the same time, careful attention should be given to the interaction between vocal characteristics and perceived personality, particularly when aiming to convey consistent and interpretable agent traits.

\section{Limitations}\label{limitations}

This study has several limitations that should be considered when interpreting the results. First, the voice stimuli were derived from a limited number of recordings selected based on self-reported personality. While this approach enabled us to ground the stimuli in participant-provided data, the extent to which these voices systematically reflect acoustic markers of introversion and extroversion was not objectively evaluated in this exploratory study and therefore cannot be determined from its findings and observations. Moreover, because each experimental condition was represented by a single synthesized voice, the observed differences cannot be unequivocally attributed to the intended personality manipulation. Instead, they may also reflect stimulus-specific vocal characteristics, such as accent, tone, speaking style, or other sociophonetic properties. For instance, a noticeable regional accent in one of the male voices may have influenced participants' perceptions. Thus, our findings should be interpreted as exploratory evidence derived from the specific voice conditions examined in this study rather than as broadly generalizable effects of vocal personality expression.

Second, we did not include an explicit acoustic-level validation of the voice stimuli (e.g., analysis of pitch, tempo, or variability). Such analyses would provide valuable insight into whether the intended personality manipulations were reflected in measurable acoustic differences across the synthesized voices. Incorporating such measures in future work would help strengthen the link between self-reported personality and its vocal expression.

Third, although the TIPI \cite{GOSLING2003504} was adapted to evaluate perceived extroversion in synthesized voices rather than self-reported personality, we did not independently evaluate the psychometric properties of this perception-based administration. Hence, while the instrument provided a structured framework for comparing perceived personality across voice conditions, future work should examine the reliability and validity of this adapted use.

Fourth, while the participant sample was sufficiently large for statistical analysis, it may not fully capture the diversity of broader populations. Participants were recruited from a single computing course at one institution, resulting in a relatively homogeneous sample that was predominantly composed of young adults and male participants. Additionally, the study focused on brief, voice-only interactions, which do not reflect the full complexity of real-world human-agent interactions that often involve multimodal cues and extended engagement. Moreover, the unequal participant distribution across gender groups should be considered when interpreting subgroup analyses, as the absence of significant effects in smaller subgroups may reflect limited statistical power rather than the absence of an underlying relationship.

Finally, the observed patterns of personality synchrony were not consistent across all participants and appeared to depend on both subgroup characteristics and interaction order. These findings point to potentially nuanced and context-dependent effects that would benefit from further investigation.

Future research can build on this work by incorporating more diverse participant populations, a broader and more rigorously controlled experimental manipulations, and more ecologically rich interaction scenarios to further examine how personality cues are perceived in human-agent interactions.

\section*{Acknowledgments}
The authors thank all individuals who voluntarily participated in this study. Their input was essential for the outcomes reported in this manuscript.

This material is based upon work supported by the National Science Foundation (NSF) under Grant No. 2434428. Any opinions, findings, or recommendations expressed in this material are those of the author(s) and do not necessarily reflect the views of the NSF.

The authors acknowledge the use of Grammarly and ChatGPT-5.5, an AI-based language model developed by OpenAI, for spell-checking, grammar, and editing assistance. Both tools were used to refine the language and enhance the manuscript's clarity. No content, figures, or images were generated by artificial intelligence (AI) for this work.

\bibliographystyle{Frontiers-Harvard}
\bibliography{sample-base}


\end{document}